# Error Probability of OSTB Codes and Capacity Analysis with Antenna Selection over Single-Antenna AF Relay Channels

S. Bogana, M. Helali, K. Paykan

*Abstract*—In this paper, the symbol error rate (SER) and the bit error rate (BER) of orthogonal space-time block codes (OSTBCs) and their achievable capacity over an amplify-and-forward (AF) relay channel with multiple antennas at source and destination and single antenna at relay node are investigated. Considered are receive antenna selection, transmit antenna selection, and joint antenna selection at both the transmitter and the receiver. The exact SERs of OSTBCs for M-PSK and square M-QAM constellations are obtained using the moment generating functions (MGFs). Also, we analyze the achievable capacity over such channels assuming antenna selection is done at the source and relay nodes. We show that a small number of selected antennas can achieve the capacity of the system in which no channel state information (CSI) is available at the source and relay nodes.

## I. INTRODUCTION

Performance of wireless communication systems can be significantly improved using diversity in space, time, or frequency [1]–[4]. Space diversity employs an array of spatially separated antennas to exploit location dependent fading characteristics in a wireless channel. This concept has been utilized in multi-input-multi-output (MIMO) systems to improve the performance in spatially uncorrelated fading environments [5]–[8]. However, as the size of a mobile terminal becomes smaller, it is more difficult to design antennas with large spatial separation. Closer antennas lead to increased correlated fading characteristics, reduces the diversity order, and therefore, lower performance improvement. Cooperative diversity [7], [9], [10] is a promising technique to overcome this limitation.

Cooperative diversity is a form of space-time diversity which makes use of multiple transmission nodes to improve received signal quality. Unlike MIMO systems, cooperative diversity relies on data transmission by several nodes. Each node acts as a virtual antenna and cooperatively transmits data to the destination. Since these nodes are located at different locations, cooperative diversity does not suffer from correlated antennas issue as much as MIMO. Relay aided cooperative wireless communication has recently been recognized as an attractive solution to improve the quality of wireless channels [17], [?], [15], [14], and [12]. Therefore, considerable attention has been paid to understanding the performance of relay channels under various conditions [6], [8], [17], [?], [12].

In this paper, the symbol error rate (SER) and the bit error rate (BER) of orthogonal space-time block codes (OSTBCs) and channel capacity with antenna selection over relay fading channel with single antenna amplify-and-forward (AF) relaying is considered. In many applications, it is desirable to keep the complexity at the relay node low. We, therefore, consider single-antenna relay node throughout this paper. In this case, the fading between each pair of the transmit and receive antennas in the presence of the relay node is characterized by double Rayleigh fading distribution, i.e., the product of two independent Rayleigh distributions [13], [16]. In order to evaluate the exact average symbol error rate (SER) of the orthogonal STBC with M-PSK and M-QAM constellations over this channel, we first derive the moment generating function (MGF) of signal-to-noise ratio (SNR) after space-time decoding. Then, using the MGF-based approach for evaluating the error performance over fading channels [18], we express the average SER of the STBC in the form of single finite-range integrals involving only the derived MGF. After error probability analysis, we consider capacity analysis for this scheme. To this aim we study the capacity of the antenna-selection, and determine the number of selected antennas (equivalently, the number of RF chains needed) so that we can match the capacity of a baseline system without antenna selection. We further prove the validity of the analytical results in the paper by means of simulation.

The remainder of this paper is organized as follows: a general model for AF relay channels with multi-antenna source and destination and single-antenna relay nodes is described in Section II and used to derive the probability distribution function (pdf) and MGF of SNR after decoding OSTBC at the destination. In Section III, antenna selection and capacity problem is discussed. Simulation results for different scenarios are considered in Section IV and Section V concludes the paper.

## II. SYSTEM MODELING

We consider a cooperative relay system with $K$ antennas at source, $N$ antennas at destination, and a single-antenna relay node as shown in Fig. 1.

### A. Relay Channel Model

It is assumed that the channel is a quasi-static flat-fading channel and full channel state information (CSI) is known at the receiver but unknown at the transmitter. The system model can then be written as

$$\mathbf{r} = \mathbf{g}\mathbf{h}^T\mathbf{x} + \mathbf{g}n_r + \mathbf{n_d} \qquad (1)$$

where **x** is the $(K \times 1)$ transmit vector, $n_r$ is the scalar noise at the relay, and $\mathbf{n_d}$ is the $(N \times 1)$ receive noise vector. **h** and **g** are the $(K \times 1)$ and $(N \times 1)$ source-relay and relay-destination channel vectors whose entries are i.i.d. zero-mean complex Gaussian random variables with variance 1/2 per dimension, respectively. Assuming independent **h** and **g**, $\mathbf{H} = \mathbf{gh}$ can be looked at as effective channel matrix whose entries are circularly symmetric complex zero-mean Gaussian random variables with variance 1/2 per dimension.

### B. pdf and MGF of the Symbol SNR after STBC Decoding

For receive antenna selection, the receiver will select $N_s$ out of $N$ antennas. If antenna selection is performed at the transmitter, it is assumed that the transmitter knows only the indices of $K_s$ out of $K$ transmit antennas sent back from the receiver via a rate-limited feedback with error-free channel. Moreover, joint antenna selection with $K_s$ transmit antennas and $N_s$ receive antenna can also be implemented. In a space-time block coded system, with $K_s$ and $N_s$ selected antennas, each block of $\tau.\kappa$ information bits is first mapped to a sequence of $\tau$ symbols $v_1, v_2, ..., v_\tau$, where each $v_i$ can be one of $M = 2^\kappa$ signal points $\{\phi_i\}_{i=1}^M$ in a M-ary constellation $\Phi$. The mapping rule $\xi$ defines the correspondence between $\kappa$ bits and a signal point in $\Phi$. The sequence $v_1, v_2, ..., v_\tau$ is then encoded into an $N_c \times N_s$ space-time block code matrix G. Here, $N_c$ is the number of rows of the space-time code matrix and the code rate is $R = \tau/N_c$ symbol per channel use. In the case of OSTBCs, the columns of G are orthogonal. The receive signals at the $N_s$ selected receive antennas over $N_c$ time periods are collectively given as:

$$\mathbf{R} = \sqrt{\lambda}\mathbf{GH} + \mathbf{Z} \tag{2}$$

Here, $\mathbf{H_s} = \mathbf{g}_{N_s}.\mathbf{h}_{K_s}{}^T$ with $\mathbf{h}_{K_s}{}^T$ and $\mathbf{g}_{N_s}$ correspond to $K_s$ and $N_s$ selected components of $\mathbf{h^T}$ and **g** , respectively. The matrix $Z$ is a $N_c \times N_s$ matrix representing additive white Gaussian noise whose entries are also CN(0, 1). The normalization factor $\lambda = \mu/N_s$ ensures that the average SNR at each receive antenna is $\mu$, independent of $N_s$. Let $\Theta$ be the squared Frobenius norm of $H$ , namely $\Theta = \|\mathbf{H_s}\|_F^2$ . The achievable SNR per symbol is given as:

$$\mu_s = \frac{\mu R}{N_s}.\Theta = \frac{\mu R}{N_s}.A.B \tag{3}$$

where $A = \|\mathbf{h}_{K_s}{}^T\|^2$ and $B = \|\mathbf{h}_{N_s}\|^2$. From (3), it is obvious that the problem of joint transmit and receive antenna selection to maximize $\mu_s$ can be decomposed into the problems of transmit antenna selection and receive antenna selection independently. In particular, $K_s$ and $N_s$ transmit and receive antennas should be selected to maximize $A$ and $B$ , respectively. In order to derive the SER and BER, this subsection first evaluates the MGF of the random variable $\Theta$. Since $A$ and $B$ are sums of $K_s$ and $N_s$ independent exponential RVs, respectively, they are central chi-square distributed with $2K_s$ and $2N_s$ degrees of freedom, $\chi^2_{2K_s}$ and $\chi^2_{2N_s}$ respectively. Following the analysis in [19], the probability density function (pdf) of the random variable $B$ is given as:

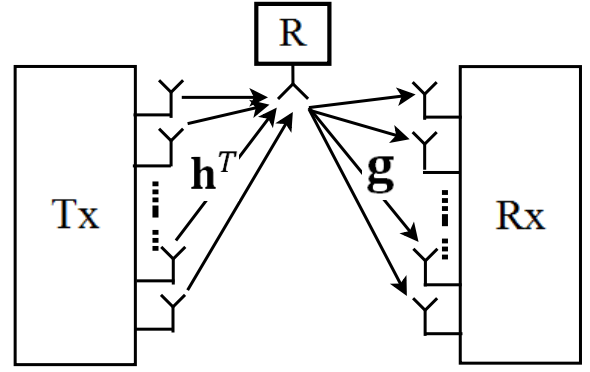

Figure 1. A MIMO cooperative scheme operating in amplify-and-forward (AF) mode with single-antenna relay node

$$p_B(b) = \binom{N}{N_s} [b^{N_s-1} \frac{e^{-b}}{\Gamma(N_s)}$$
$$+ \sum_{i=1}^{N-N_s} \chi_i e^{-b} \left( e^{-\frac{ib}{N_s}} - \sum_{k=1}^{N_s-1} \gamma_{i,k} b^{k-1} \right)] \tag{4}$$

where

$$\chi_i = (-1)^i \binom{N-N_s}{i} \left(\frac{-N_s}{i}\right)^{N_s-1} \tag{5}$$

and

$$\gamma_{i,k} = \frac{1}{\Gamma(k)} \left(\frac{-i}{N_s}\right) \tag{6}$$

The pdf of $A$ is also expressed in a similar form. The pdf of $\Theta$ is, therefore, given by

$$p_\Theta(\theta) = \int_0^\infty p_B(v) p_A(\frac{\theta}{v}) \frac{1}{v} dv \tag{7}$$

The moment generating function of $\Theta$ with joint receive and transmit antenna selection can then be computed as

$$\phi_\Theta^s(s) = E\left\{e^{-s.\Theta}\right\} = \int_0^\infty p_\Theta(\theta) e^{-s\theta} d\theta$$

As shown in Appendix, the pdf of $\Theta$ , $p_\Theta(\theta)$, and the MGF of $\Theta$ ,$\phi_\Theta^s(s)$ , can be written in closed-form expressions. In particular, from (22) it can be seen that the MGF of $\Theta$ is expressed in a closed form as the sum of various functions, each having the same form of $f(s; x_1, x_2, x_3, x_4) = x_1 s^{x_2} U(x_2, x_3, x_4/s)$, where $U(a, b, x)$is the confluent hyper-geometric function of the second kind [?] and $x_1, x_2, x_3, x_4$ are some constants. For example, in the case of receive antenna selection only, since $K - K_s = 0$ and the fact that $\chi_0 = 0$ and $\gamma_{i,0} = 0$, the MGF of $\Theta$ in (22) can be written in a more compact form as:

$$\phi_{\Theta}^s(s) = \frac{2 \begin{pmatrix} N \\ N_s \end{pmatrix}}{\Gamma(K)} \left( \frac{\Gamma(K)}{2} s^{-N_s} U\left(N_s, N_s - K + 1, \frac{1}{s}\right) \right.$$
$$+ \sum_{k=1}^{N-N_s} \chi_k \left\{ \frac{\Gamma(N)}{2} \left(\frac{N_s + k}{N_s}\right)^{K-1} s^{-K} U\left(K, K\frac{N_s + k}{N_s s}\right) \right.$$
$$\left. \left. - \sum_{l=1}^{N_s-1} \gamma_{k,l} \frac{\Gamma(l)\Gamma(N)}{2} s^{-l} U\left(l, 1 + l - K, \frac{1}{s}\right) \right\} \right) \tag{8}$$

With the closed-form expressions of the MGF of $\Theta$ presented in (8), the exact SER of OSTBCs for standard M-PSK and square M-QAM constellations can be computed straightforwardly. More specifically, by averaging the conditional SERs for M-PSK and square M-QAM constellations given in [19] over the pdf of the instantaneous SNR $\Theta$, one obtains the following expressions for the general case of joint receive and transmit antenna selection:

$$P_{\text{symbol}}^{MPSK}(\text{error}) = \frac{1}{\pi} \int_0^{\pi - \pi/K} \phi_{\Theta}^s \left( \frac{\frac{\mu R}{N_s} g_{MPSK}}{\sin^2\theta} \right) \mathrm{d}\theta \tag{9}$$

$$P_{\text{symbol}}^{MQAM}(\text{error}) = \frac{4q}{\pi} \int_0^{\frac{\pi}{2}} \phi_{\Theta}^s \left( \frac{\frac{\mu R}{N_s} g_{MQAM}}{\sin^2\theta} \right) \mathrm{d}\theta$$
$$- \frac{4q^2}{\pi} \int_0^{\frac{\pi}{4}} \phi_{\Theta}^s \left( \frac{\frac{\mu R}{N_s} g_{MQAM}}{\sin^2\theta} \right) \mathrm{d}\theta \tag{10}$$

where $\phi_{\Theta}^s(\theta)$ is the MGF of $\Theta$, given in (8) and $q = (1 - 1/\sqrt{M})$. It can be seen that the single integrals in (9) and (10) are over finite ranges. Therefore, the exact SERs for M-PSK and square M-QAM can be computed easily. This fact holds for transmit antenna selection and receive antenna selection as well, since the MGFs of $\phi_{\Theta}^s(\theta)$ in all three cases have similar expressions.

## III. Antenna Selection and Capacity

In the antenna-selection problem, we assume that the transmitter (also the receiver) selects one or more of the available antennas. Beyond this, the transmitter does not have any additional information, thus the power will be equally distributed among the selected antennas. Assuming equal power splitting among antennas, the capacity of the channel (without any antenna selection) is

$$C = \log\left(\det\left[\mathbf{I}_N + \frac{\rho}{K}\mathbf{H^T H}\right]\right) = \log\left(1 + \frac{\rho}{K}\|\mathbf{h}\|^2\|\mathbf{g}\|^2\right) \tag{11}$$

thus, the instantaneous SNR, normalized by the received SNR, can be defined as

$$\delta \triangleq \frac{\|\mathbf{h}\|^2\|\mathbf{g}\|^2}{K} \tag{12}$$

and the capacity (conditioned on ) is $C = \log(1 + \rho\delta)$, which is a monotonic function of SNR. Since the random variables $\|\mathbf{h}\|^2$ and $\|\mathbf{g}\|^2$ are distributed chi-square $\chi_{2K}^2$ and $\chi_{2N}^2$, respectively, the average normalized SNR for the channel is $E\{\delta\} = N$ Now considering the channel with selection, the selected (sub)channel can be represented as $\mathbf{H}_{N_s \times K_s}^s = \mathbf{g}_s \cdot \mathbf{h}_s{}^T$, where $\mathbf{g}_s$ and $\mathbf{h}_s$ are the channel vectors of selected antennas at the transmit and receive sides. Power is split equally between selected transmit antennas. The normalized instantaneous SNR for the selected channel is

$$\delta^s \triangleq \frac{\|\mathbf{h}_s\|^2\|\mathbf{g}_s\|^2}{K_s} \tag{13}$$

and the capacity formula is $C = \log(1 + \rho\delta^s)$. In order to maximize the capacity subject to selection, it suffices to maximize $\delta^s$. The average per-element energy of a vector is less than the energy of the largest vector element, that is

$$\frac{\|\mathbf{h}_s\|^2}{K_s} \leq \max_i |\mathbf{h}_s(i)|^2 \tag{14}$$

Therefore the best strategy is to select only one antenna at the transmitter, the one with the highest channel gain. Transmit antenna selection is possible with minimal feedback of $\lceil \log K \rceil$ bits. For $K_s = 1$ and $N_s = N$, it can be easily shown that

$$E\{\delta^s\} = N\sum_{k=1}^K \frac{1}{k} \tag{15}$$

We now proceed to the receive-side selection. The diversity obtained by selecting out of antennas is known as generalized selection diversity [21], and has been extensively studied in the literature. Using results from [21] we can calculate the average normalized SNR for antenna selection with $K_s = 1$ and arbitrary $N_s$

$$E\{\delta^s\} = \sum_{k=1}^K \frac{1}{k}\left(N_s + N_s\sum_{i=N_s+1}^N \frac{1}{i}\right) = \Delta(N_s) \tag{16}$$

As mentioned earlier, we wish to find the minimum number of selected receive antennas such that the equivalent SNR is no less than $E\{\delta\}$. Considering $\Delta(N_s)$ that above is a monotonic function, the number of selected antennas will be

$$\hat{N}_s = \min \Delta^{-1}(\delta^s) \quad s.t. \quad \delta^s \geq \delta \tag{17}$$

## IV. Simulation and Discussion

We provide the results of our analysis and compare them with simulation results in order to verify the analysis. For two transmit antennas, we use the following one-rate STBC (Alamouti code) given in [11].

$$g_2 = \begin{pmatrix} x_1 & x_2 \\ -x_2^* & x_1^* \end{pmatrix} \tag{18}$$

where the superscript denotes the complex conjugate. Fig. 2. shows the simulation results and the exact analysis of the SER versus average symbol SNR per receive antenna for the STBCs and with 8-PSK in that we use antenna selection with $K_s = N_s = 2$. Fig. 3. shows the simulation results and the exact analysis of the BER for 16-QAM in that we use antenna selection with $K_s = 2$ and $N_s = 1$. In fig. 2. we depicted the curves for two different case one without antenna selection and the other with joint antenna selection in transmitter and receiver sides to compare the result. In fig. 3. we consider joint antenna selection and antenna selection in receiver side.

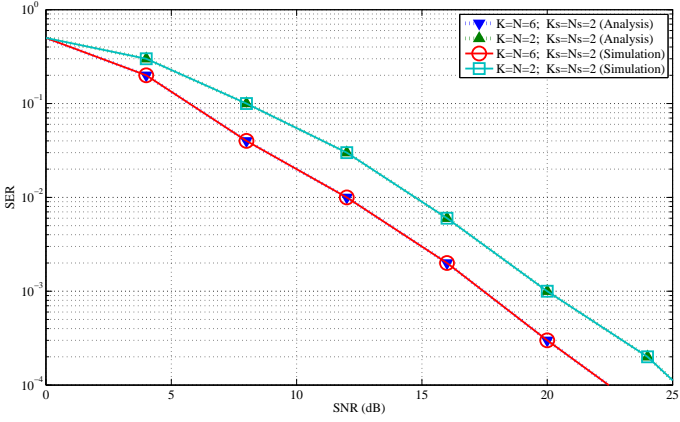

Figure 2. SERs versus average SNR per receive antenna for the STBC $g_2$ with 8-PSK

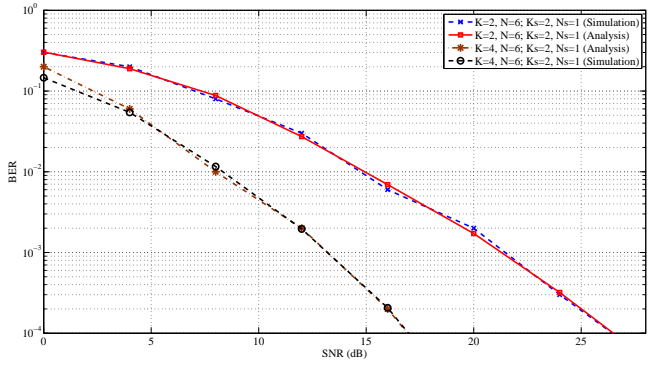

Figure 3. BERs versus average SNR per receive antenna for the STBC $g_2$ with 16-QAM

Table I shows the calculated values of $N_s$ for various systems. It is interesting to observe that across a large group of systems, a small number of selected antennas is sufficient to match the capacity of a baseline system with no transmit CSI, but with full hardware at both sides. Antenna selection requires a small amount of feedback, but has dramatically smaller hardware requirements.

Table I
TABLE SHOWS THE NUMBER OF RECEIVE ANTENNAS SELECTED TO MATCH CAPACITY OF BASELINE SYSTEM

| | Rx Antennas | | | | | | | | |
|---|---|---|---|---|---|---|---|---|---|
| Tx Antennas | 2 | 3 | 4 | 5 | 6 | 7 | 8 | 9 | 10 |
| 2 | 1 | 1 | 1 | 2 | 2 | 2 | 2 | 3 | 3 |
| 3 | 1 | 1 | 1 | 1 | 2 | 2 | 2 | 2 | 2 |
| 4 | 1 | 1 | 1 | 1 | 1 | 1 | 1 | 2 | 2 |
| 5 | 1 | 1 | 1 | 1 | 1 | 1 | 1 | 1 | 2 |
| 6 | 1 | 1 | 1 | 1 | 1 | 1 | 1 | 1 | 1 |
| 7 | 1 | 1 | 1 | 1 | 1 | 1 | 1 | 1 | 1 |

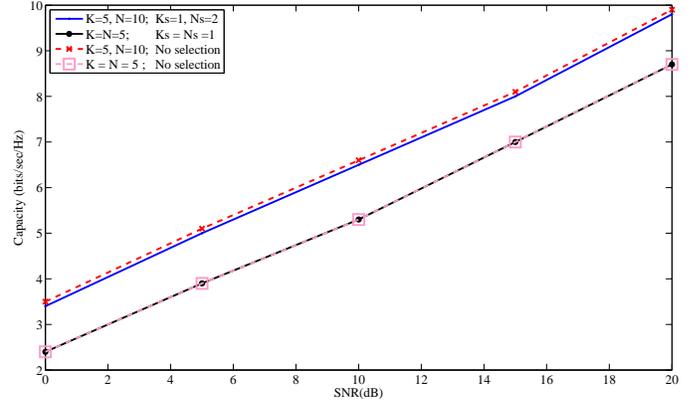

Figure 4. Ergodic channel capacity with and without antenna selection.

Fig. 4. shows the ergodic capacity of the channel for two cases: $K = N = 5$, out of which a $1 \times 1$ system is chosen; and $K = 5, N = 10$, out of which a $1 \times 2$ system is chosen. Capacity is compared between the antenna selection and the full system. The Monte Carlo simulation is performed over 1000 independent channel realizations. The simulation results show that at the cost of a few bits of feedback, channel in this scheme can be reduced to a low-order SIMO channel without a considerable loss in the capacity.

## V. CONCLUSION

This paper examined the SER and BER of OSTBCs over single antenna relay channel with antenna selection and capacity analysis of this structure. For standard M-PSK and square M-QAM constellations, closed-form expressions were provided to compute exactly the SER. We analysed the capacity of antenna selection over this scheme and showed that only a very small number of antennas need be selected (only one at transmitter, and very few at receiver) to match the performance of a baseline system that operates with all antennas.

## APPENDIX A
## THE PDF AND MGF OF $\Theta$ FOR JOINT RECEIVE AND TRANSMIT ANTENNA SELECTION

Using (7) and based on the expression of the modified Bessel function of the second kind, namely

$$K_\nu(x_1.x_2) = \frac{x_2}{2} \int_0^\infty e^{-\frac{x_1}{2}\left(r + \frac{x_2^2}{r}\right)} r^{(1-\nu)} dr \qquad (19)$$

the pdf of $\Theta$ for the general case of joint receive and transmit antenna selection is given as:

$$\qquad (20)$$

By using [20] one obtains the following integral regarding

$$\frac{p_\Theta(\theta)}{2\binom{K}{K_s}\binom{N}{N_s}} = \frac{\theta^{\frac{(K_s+N_s)}{2}-1}}{\Gamma(K_s)\Gamma(N_s)}\mathrm{K}_{N_s-K_s}\left(2\sqrt{\theta}\right)$$

$$+ \frac{1}{\Gamma(N_s)}\sum_{k=0}^{K-N_s}\chi_k\left[\left(\frac{K_s+k}{K_s}\theta\right)^{\frac{N_s-1}{2}}\mathrm{K}_{N_s-1}\left(2\sqrt{\theta\frac{K_s+k}{K_s}}\right)\right.$$

$$- \sum_{p=0}^{K_s-1}\gamma_{k,p}\theta^{\frac{p+N_s}{2}-1}\mathrm{K}_{p-N_s}\left(2\sqrt{\theta}\right)\Bigg]$$

$$+ \frac{1}{\Gamma(K_s)}\sum_{i=0}^{N-N_s}\chi_i\left[\left(\frac{N_s+i}{N_s}\theta\right)^{\frac{K_s-1}{2}}\mathrm{K}_{K_s-1}\left(2\sqrt{\theta\frac{N_s+i}{N_s}}\right)\right.$$

$$- \sum_{j=0}^{N_s-1}\gamma_{i,j}\theta^{\frac{j+K_s}{2}-1}\mathrm{K}_{j-K_s}(2\sqrt{\theta})\Bigg]$$

$$+ \sum_{i=0}^{N-N_s}\sum_{k=0}^{K-K_s}\chi_i\chi_k\mathrm{K}_0\left(2\sqrt{\theta\left(\frac{K_s+k}{K_s}\right)\left(\frac{N_s+i}{N_s}\right)}\right)$$

$$- \left(\sum_{i=0}^{N-N_s}\sum_{k=0}^{K-K_s}\sum_{p=0}^{N_s-1}\chi_i\chi_k\gamma_{k,p}\left(\frac{N_s+i}{N_s}\theta\right)^{\frac{p-1}{2}}\right.$$

$$\left.\cdot\,\mathrm{K}_{p-1}\left(2\sqrt{\theta\frac{N_s+i}{N_s}}\right)\right)$$

$$- \left(\sum_{k=0}^{K-K_s}\sum_{i=0}^{N-N_s}\sum_{j=0}^{N_s-1}\chi_k\chi_i\gamma_{i,j}\left(\frac{K_s+k}{K_s}\theta\right)^{\frac{j-1}{2}}\right.$$

$$\left.\cdot\,\mathrm{K}_{j-1}\left(2\sqrt{\theta\frac{K_s+k}{K_s}}\right)\right)$$

$$+ \sum_{i=0}^{N-N_s}\sum_{j=0}^{N_s-1}\sum_{k=0}^{K-K_s}\sum_{p=0}^{K_s-1}\chi_i\gamma_{i,j}\chi_k t_{k,p}\theta^{\frac{p+j}{2}-1}\mathrm{K}_{p-j}\left(2\sqrt{\theta}\right)$$

$$\frac{\phi_\Theta^s(s)}{\binom{N}{N_s}\binom{M}{K_s}} = s^{-N_s}U\left(N_s,N_s-K_s+1,\frac{1}{s}\right)$$

$$+ \sum_{k=0}^{K-K_s}\chi_k\left[\left(\frac{K_s+k}{K_s}\right)^{N_s-1}s^{-N_s}U\left(N_s,N_s,\frac{K_s+k}{K_s s}\right)\right.$$

$$- \sum_{p=0}^{K_s-1}\gamma_{k,p}\Gamma(p)s^{-p}U\left(p,p+1-N_s,\frac{1}{s}\right)\Bigg]$$

$$+ \sum_{i=0}^{N-N_s}\chi_i\left[\left(\frac{N_s+i}{N_s}\right)^{K_s-1}s^{-K_s}U\left(K_s,K_s,\frac{N_s+i}{N_s s}\right)\right.$$

$$- \sum_{j=0}^{N_s-1}\gamma_{i,j}(j)s^{-j}U\left(j,1+j-K_s,\frac{1}{s}\right)\Bigg]$$

$$+ \sum_{i=0}^{N-N_s}\sum_{k=0}^{K-K_s}\chi_i\chi_k s^{-1}U\left(1,1,\frac{(N_s+i)(K_s+k)}{K_s N_s s}\right)$$

$$- \left(\sum_{i=0}^{N-N_s}\sum_{k=0}^{K-K_s}\sum_{p=0}^{N_s-1}\chi_i\chi_k\gamma_{k,p}\Gamma(p)\left(\frac{N_s+i}{N_s}\right)^{p-1}s^{-p}\right.$$

$$\left.U\left(p,p,\frac{N_s+i}{N_s s}\right)\right)$$

$$- \left(\sum_{k=0}^{K-K_s}\sum_{i=0}^{N-N_s}\sum_{j=0}^{N_s-1}\chi_k\chi_i\gamma_{i,j}\Gamma(j)\left(\frac{K_s+k}{K_s}\right)^{j-1}s^{-j}\right.$$

$$\left.U\left(j,j,\frac{K_s+k}{K_s s}\right)\right)$$

$$+ \left(\sum_{i=0}^{N-N_s}\sum_{j=0}^{N_s-1}\sum_{k=0}^{K-K_s}\sum_{p=0}^{K_s-1}\chi_i\gamma_{i,j}\chi_k\gamma_{k,p}\Gamma(j)\Gamma(p)s^{-p}\right.$$

$$\left.U\left(p,1+p-j,\frac{1}{s}\right)\right)$$

the modified Bessel function of the second kind:

$$\int_0^\infty \theta^{\mu-\frac{1}{2}}e^{-s\theta}\mathrm{K}_{2\nu}\left(2\chi\sqrt{\theta}\right)d\theta =$$

$$\frac{1}{2}\Gamma\left(\mu+\nu+\frac{1}{2}\right)\Gamma\left(\mu-\nu+\frac{1}{2}\right)$$

$$.\chi^{2\nu}s^{-(\mu+\nu+\frac{1}{2})}U\left(\mu+\nu+\frac{1}{2},1+2\nu,\frac{\chi^2}{s}\right) \quad (21)$$

From (20) and (21) and after some manipulations, the MGF of $\Theta$ can be computed as follows:

$$(22)$$